\documentclass[aps,preprint]{revtex4}
\usepackage [dvips]{graphicx}
\def\beq{\begin{equation}}
\def\eeq{\end{equation}}
\def\pmb{\mathrm}
\begin{document}

\title{The cluster structure of the inner crust of neutron stars in the
Hartree-Fock-Bogoliubov approach}

\author{F. Grill$^{a,b}$,
J. Margueron$^{c}$, N. Sandulescu$^{d}$\footnote{corresponding
author (email:sandulescu@theory.nipne.ro)} }
\affiliation { $^{a}$
Dipartimento di Fisica, Universit\'a degli Studi di Milano,
Via Celoria 16, 20133 Milan, Italy \\
$^b$ Centro de F\'{i}sica Computacional, Department of Physics, University of Coimbra, PT-3004-516 Coimbra, Portugal \\
$^c$ Institut de Physique Nucleaire, Universit\'e Paris-Sud, Orsay Cedex, France \\
$^{d}$ National Institute of Physics and Nuclear Engineering,
76900, Bucharest, Romania}

\begin{abstract}
We analyse how the structure of the inner curst is influenced by the pairing
correlations. The inner-crust matter, formed by nuclear clusters immersed
in a superfluid neutron gas and ultra-relativistic electrons, is treated
in the Wigner-Seitz approximation. The properties of the Wigner-Seitz cells, i.e.,
their neutron to proton ratio and their radius at a given baryonic
density, are obtained from the energy minimization at beta
equilibrium. To obtain the binding energy of baryonic matter we
perform Skyrme-HFB calculations with zero-range
density-dependent pairing forces of various intensities. We find
that the  Wigner-Seitz cells have much smaller numbers of protons 
compared to previous calculations. For the dense cells the binding
energy of the configurations with small proton numbers do not
converge to a well-defined minimum value which precludes the
determination of their structure. We show that for these cells
there is a significant underestimation of the binding energy due to the
boundary conditions at the border of the cells imposed through the
Wigner-Seitz approximation.

\end{abstract}

\maketitle

\section{Introduction}

The inner-crust of neutron stars extends from the so-called neutron drip density,
$\rho_d \approx 4 \times 10^{11}$ g cm$^{-3}$, defined as the density where the
neutrons start to drip out from the nuclei of the crust, up to a density of about
$\rho \approx 1.4 \times 10^{14}$ g cm$^{-3}$ at which there is a transition
towards the uniform core matter. The size and the precise density limits of 
the inner crust depend on the star mass and on the equation of state
employed in the star models
\cite{glendenning,haensel_book}.

The inner-crust matter of non-accreting cold neutron stars is most probably formed
by a crystal lattice of nuclear clusters immersed in a sea of low-density
superfluid neutrons and ultra-relativisitic electrons. It is generally
considered that in most of its part the inner-crust is formed of
nuclei-like clusters. More complex "pasta" structures (e.g., rods,
plates, bubbles) are expected to be formed in the transition region
between the inner crust and the core matter,
see for instance Refs.~\cite{glendenning,haensel_book} and references therein.

\begin{figure}[ht]
\begin{center}
\includegraphics[scale=1,angle=-90]{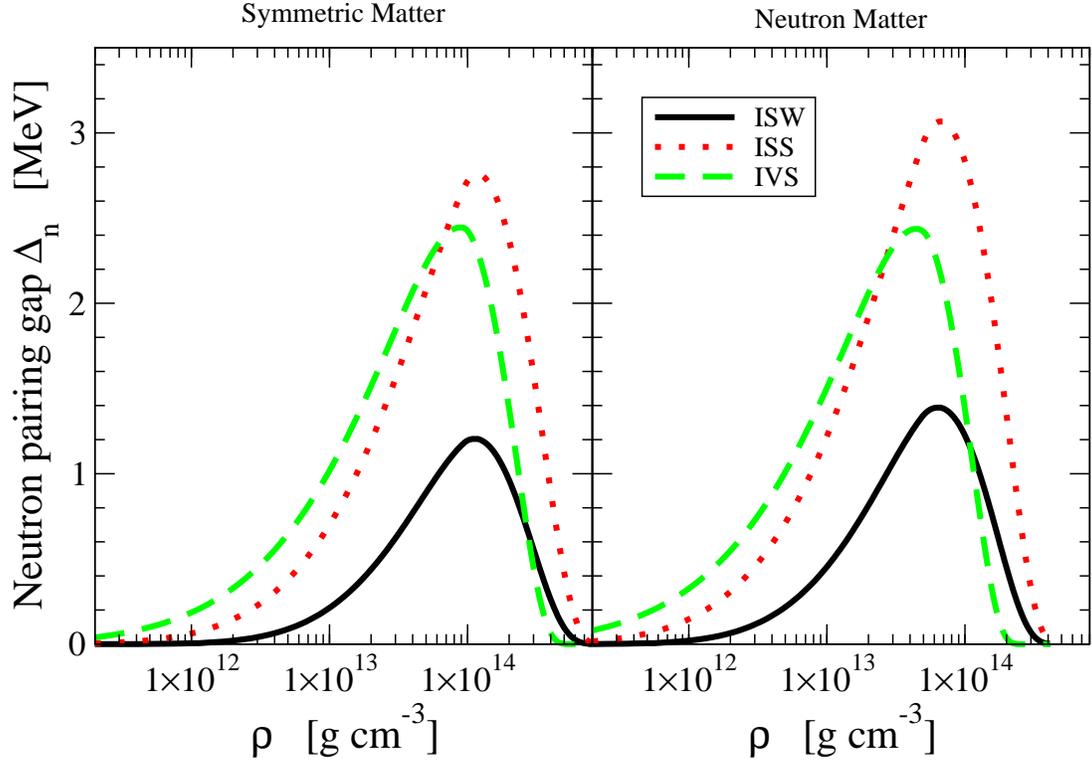}
\caption{(color online) Neutron pairing gap for the interactions
ISW (isoscalar weak), ISS (isoscalar strong)
and IVS (isovector-isoscalar)
in symmetric nuclear matter and in neutron matter.}
\label{fig:gap}
\end{center}
\end{figure}

The first microscopic calculation of the inner-crust structure, still
used as a benchmark in neutron stars studies (e.g., Refs.~\cite{page,pizzochero,monrozeau})
was performed by Negele and Vautherin in 1973~\cite{negele}.
In this work the crystal lattice is divided in spherical cells which are treated in the
Wigner-Seitz (WS) approximation.
The nuclear matter from each cell is described in the framework of Hartree-Fock (HF)
approximation based on the Density Matrix Expansion (DME)~\cite{negele72}.
This approach was preferred to the Density-Dependent Hartree-Fock theory~\cite{negele70}
in order to reduce computational complication induced by the non-local exchange
potential.
The parameters of the DME theory were adjusted to reproduce the experimental
binding energies of atomic nuclei and the theoretical calculation of infinite neutron
matter available at that time.
The spin-orbit interaction was taken into account for the protons but neglected for the neutrons.
The HF equations were solved in coordinate representation imposing mixed Dirichlet-Neuman
boundary conditions at the border of the cells.
The properties of the WS cells found in Ref.~\cite{negele}, determined for a
limited set of densities, are shown in Table~\ref{tab:NV}.
The most remarkable result of this calculation is that the majority of the cells
have semi-magic and magic proton numbers, i.e., Z=40,50.
This indicates that in these calculations there are strong proton shell effects,
as in isolated atomic nuclei.
However, as seen from Fig.~\ref{fig:gap} of Ref.~\cite{negele} the  energies corresponding
to the cells configurations based on various Z numbers are in fact very close to each other.
This fact raises several questions:
i) what is the sensitivity of the results on the nuclear interaction,
ii) how much the  pairing correlations influence the cells structure,
iii) how reliable is the WS approximation.

The effect of pairing correlations on the structure of WS cells was investigated
in Ref.~\cite{baldo} within the HF+BCS approach.
In the most recent version of these calculations the authors solved the HF+BCS equations
with a mixture between the phenomenological functional of Fayans et al~\cite{fayans},
employed in the nuclear cluster region, and a microscopical functional derived
from Bruckner-Hartree-Fock calculations in infinite neutron matter.
The latter was used to describe the neutron gas in the WS cells.
In this framework it was found that the cells have not a magic or semi-magic
number of protons, as in Ref.~\cite{negele}. It was also found that pairing can
change significantly the structure of the cells compared to HF calculations.
These findings show that in order to determine the most probable structure of the inner
crust one needs more investigations based on various effective interactions and
many-body approximations.

In this study we analyse the effect of pairing on inner crust structure
in the Hartree-Fock-Bogoliubov approach (HFB). This approach offers better
grounds than HF+BCS approximation for treating pairing correlations
in non-uniform nuclear matter with both bound and unbound neutrons.
To investigate the dependence of the inner crust structure on pairing,
the HFB calculations are performed with three different density-dependent pairing
forces adjusted to reproduce various pairing scenarios in nuclear matter.

In principle, the symmetries of the inner crust lattice should be taken properly
into account when inner crust structure is determined. Since imposing the exact
lattice symmetries in microscopic self-consistent models is a very difficult
task (for approximative solutions to this problem see
~\cite{chamel2010} and the references therein) we solve the HFB
equations in the WS
approximation, as commonly done in
 inner crust studies \cite{negele,baldo}. This approximation induces,
through the boundary conditions at the border of the cells, an artificial
shell structure in the  energy spectra of nonlocalized  neutrons \cite{baldo2006}.
The errors caused by the spurious shells, which affect mainly the high
density cells, are estimated  by using the method proposed in
Ref.\cite{margueron2007}.

 The paper is organized as follows: in Section II we describe
 how the cells energy is calculated in the HFB and WS
 approximations, in Section III we present the equations for beta equilibrium
 and in Section IV we discuss the results of the calculations.

\section{The energy of Wigner-Seitz cells}

As in Ref.~\cite{negele}, the lattice structure of the inner crust is described as a set
of independent cells of spherical symmetry treated in the WS approximation.
For baryonic densities below $\rho \approx 1.4 10^{14} g/cm^3 $, each cell has in its center a
nuclear cluster (bound protons and neutrons) surrounded by low-density and delocalized
neutrons and immersed in a uniform gas of ultra-relativistic electrons which assure
the charge neutrality. At a given baryonic density the structure of the cell,
i.e., the N/Z ratio and the cell radius is determined from the minimization over N and Z
of the total energy under the condition of beta equilibrium.
The energy of the cell, relevant for determining the cell structure, has contributions from
the nuclear and the Coulomb interactions. Its expression is written in the following form
\begin{equation}
E = E_M + E_N+T_e+E_L .
\label{eq:energy}
\end{equation}
The first term is the mass difference $E_M=Z(m_p+m_e)+(N-A)m_n$
where N and Z are the number of neutrons and protons in the cell
and A=N+Z. $E_N$ is the binding energy of the nucleons, which
includes the contribution of proton-proton Coulomb interaction
inside the nuclear cluster. $T_e$ is the kinetic energy of the
electrons while $E_L$ is the lattice energy which takes into
account the electron-electron and electron-proton interactions.
The contribution to the total energy coming from the interaction
between the WS cells \cite{oyamatsu} it is not considered since
it is very small compared to the other terms of Eq.(1).
Notice also that the gravitational energy is not taken into 
account in the energy minimization because its variation at
the nuclear scale is negligible.

\subsection{The nuclear binding energy in the HFB approach}

In the present study the nuclear binding energy of the WS cells is calculated
in the framework of HFB approach. For the particle-hole channel we use a Skyrme-type
 interaction of the standard form \cite{sly4}, i.e.,
\begin{eqnarray}
\label{eq:skyrme}
V_{\rm Skyrme}(\pmb{r_i}, \pmb{r_j})  &=&  t_0(1+x_0 P_\sigma)\delta({\pmb{r}_{ij}}) \nonumber \\
& &+\; \frac{1}{2} t_1(1+x_1 P_\sigma)\frac{1}{\hbar^2}\left[p_{ij}^2\,\delta({\pmb{r}_{ij}})
+\delta({\pmb{r}_{ij}})\, p_{ij}^2 \right]\nonumber\\
& &+\; t_2(1+x_2 P_\sigma)\frac{1}{\hbar^2}\pmb{p}_{ij}.\delta(\pmb{r}_{ij})\,
 \pmb{p}_{ij} \nonumber \\
& &+\; \frac{1}{6}t_3(1+x_3 P_\sigma)\rho(\pmb{r})^\gamma\,\delta(\pmb{r}_{ij})
\nonumber\\
& &+\; \frac{\rm i}{\hbar^2}W_0(\mbox{\boldmath$\sigma_i+\sigma_j$})\cdot
\pmb{p}_{ij}\times\delta(\pmb{r}_{ij})\,\pmb{p}_{ij}  \quad ,
\end{eqnarray}
where $\pmb{r}_{ij} = \pmb{r}_i - \pmb{r}_j$, $\pmb{r} = (\pmb{r}_i +
\pmb{r}_j)/2$, $\pmb{p}_{ij} = - {\rm i}\hbar(\pmb{\nabla}_i-\pmb{\nabla}_j)/2$
is the relative momentum, and $P_\sigma$ is the two-body
spin-exchange operator. The parameters of the force we have used in this study correspond
to the Skyrme force SLy4 \cite{sly4}. This force is often employed to describe both atomic nuclei
and neutron stars properties.

The pairing correlations are described with a zero range density dependent interaction
of the following type:
\begin{equation}
V_{\rm Pair}(\pmb{r_i}, \pmb{r_j})= V_0 \; g_{\rm Pair}[\rho_n(\pmb{r}),\rho_p(\pmb{r})] \; \delta(\pmb{r}_{ij})\, ,
\label{eq:vpair}
\end{equation}
where $g_{\rm Pair}[\rho_n,\rho_p]$ is a functional of neutron and proton densities.
In the calculations we use two different functionals for $g_{\rm Pair}[\rho_n(\pmb{r}),\rho_p(\pmb{r})]$.
The first one, called below isoscalar (IS) pairing force, depends  only on the total
particle density, $\rho(r)=\rho_n(r)+\rho_p(r)$. Its expression is given by
\begin{equation}
g_{\rm Pair}[\rho_n(\pmb{r}),\rho_p(\pmb{r})]= 1-\eta \left(\frac{\rho(r)}{\rho_0}\right)^\alpha \; ,
\label{eq:is}
\end{equation}
where $\rho_0$ is the saturation density of the nuclear matter.
This effective pairing interaction is extensively used in nuclear structure
calculations and it was also employed for describing
pairing correlations in the inner crust of neutron stars
\cite{sandulescu1,sandulescu2,ns,monrozeau}.
The parameters are chosen to reproduce in infinite neutron matter two possible pairing
scenarios \cite{lombardo}, corresponding to a maximum gap of about 3 MeV (strong pairing scenario,
hereafter named ISS) and, respectively, to a maximum gap around 1 MeV
(weak pairing scenario, called below ISW). These two scenarios are simulated by
 two values of the pairing strength, i.e., V$_0$=\{-570,-430\} MeV fm$^{-3}$.
The other parameters are taken the same for the strong and the weak pairing,
i.e., $\alpha$=0.45, $\eta$=0.7 and $\rho_0$=0.16 fm$^{-3}$.
The energy cut-off in the pairing tensor (8), necessary to cure the divergence
associated to the zero range of the pairing force, was introduced through
 the factor $e^{- E_i/100}$ acting for $E_i > 20$ MeV,
where $E_i$ are the HFB quasiparticle energies.

The second pairing functional, referred below as isovector-isoscalar (IVS) pairing, depends
explicitly on neutron and proton densities and has the following form \cite{margueron2008}
\begin{equation}
g_{\rm Pair}[\rho_n(\pmb{r}),\rho_p(\pmb{r})]= 1 - \eta_s (1-I(r)) \left(\frac{\rho(r)}{\rho_0}\right)^{\alpha_s}
-\eta_n I(r) \left(\frac{\rho(r)}{\rho_0}\right)^{\alpha_n} \; ,
\label{eq:is+iv}
\end{equation}
where $I(r)=\rho_n(r)-\rho_p(r)$. As shown in Ref. ~\cite{margueron2008}, this pairing functional
describes well the two-neutron separation energies and the odd-even mass differences
in semi-magic nuclei. In the present calculations for this pairing functional we have used the
parameters  V$_0$= -703.86 MeV fm$^{-3}$, $\eta_s$=0.7115, $\alpha_s$=0.3865, $\eta_n$=0.9727,
$\alpha_n$=0.3906, with the same cut-off prescription as for the IS pairing forces.

The pairing gaps in symmetric matter and neutron matter predicted by the three pairing forces
introduced above are represented in Fig.~\ref{fig:gap} for a wide range of sub-nuclear 
densities. It can be seen that the IVS force gives a maximum gap closer to the strong
isoscalar force.

For the zero range interactions introduced above and for spherically symmetric systems
the radial HFB equations are given by
\begin{equation}
\begin{array}{c}
\left( \begin{array}{cc}
h(r) - \mu & \Delta (r) \\
\Delta (r) & -h(r) + \mu
\end{array} \right)
\left( \begin{array}{c} U_i (r) \\
 V_i (r) \end{array} \right) = E_i
\left( \begin{array}{c} U_i (r) \\
 V_i (r) \end{array} \right) ~,
\end{array}
\label{eq:hfb}
\end{equation}
\\
where $U_i$, $V_i$ are the upper and lower components of the
radial HFB wave functions, $\mu$ is the chemical potential while
$h(r)$ and $\Delta(r)$ are the mean field Hamiltonian and pairing field,
respectively. They depend on  particle density $\rho(r)$, abnormal pairing
tensor $\kappa(r)$, kinetic energy density $\tau(r)$ and spin
density $J(r)$ defined by:
\begin{equation}
\rho(r) =\frac{1}{4\pi} \sum_{i} (2j_i+1) V_i^* (r)
V_i (r)
\label{eq:density}
\end{equation}
\begin{equation}
\kappa(r) = \frac{1}{4\pi} \sum_{i} (2j_i+1) U_i^* (r)
V_i (r)
\label{eq:kappa}
\end{equation}
\beq
J(r) = \frac{1}{4\pi} \sum_i (2j_i+1)
[j_i(j_i+1)-l_i(l_i+1)-\frac{3}{4}]
\ V_i^2
\label{eq:j}
\eeq
\beq
\tau(r) = \frac{1}{4\pi} \sum_{i} (2j_i+1)
[(\frac{dV_i}{dr}-\frac{V_i}{r})^2 +\frac{l_i(l_i+1)}{r^2} V_i^2 ]
\eeq
\\
\noindent
The general expressions of the mean field in terms of the densities are given
in Ref.~\cite{dobaczewski}. The pairing field has a simple form, i.e.,
\begin{equation}
\Delta(r)= \frac{1}{2} \; g_\mathrm{Pair}[\rho_n(r),\rho_p(r)] \; \kappa(r).
\end{equation}

 The HFB equations are solved in coordinate space and imposing the
 following boundary conditions at the border of the WS
cells~\cite{negele}: i) even parity wave functions vanish at $r=R_{WS}$;
ii) first derivatives of odd-parity wave functions vanish at $r=R_{WS}$.
With these mixed boundary conditions at the cell border the continuous
quasiparticle spectrum of the unbound neutrons is discretized.

Compared to the usual HFB calculations done for nuclei,
in a WS cell the mean field of the protons has an additional
contribution coming from the interaction of the protons with the electrons.
Thus, the total proton mean field in the cell is given by
\begin{equation}
u^p(r)=u_\mathrm{nucl}^{pp}(r)+u_\mathrm{Coul}^{pp}(r)+u_\mathrm{Coul}^{pe}(r),
\end{equation}
where $u_\mathrm{nucl}^{pp}(r)$ is the nuclear part of the mean field, given by
the Skyrme interaction, while $u_\mathrm{Coul}^{pp}$
and $u_\mathrm{Coul}^{pe}$ are the mean fields corresponding to the Coulomb
proton-proton and proton-electron interactions. The proton-proton Coulomb mean field
has the standard form
\begin{equation}
u_\mathrm{Coul}^{pp}(r)=e^2\int \! d^3r' \, \rho_p(r')\frac{1}{|r-r'|}-
e^2\left(\frac{3}{\pi}\rho_p(r)\right)^{1/3}~ ,
\end{equation}
where the first and the second terms correspond, respectively, to the direct and
the exchange part of proton-proton Coulomb interaction. The latter is evaluated
in the Slater approximation.

The mean field corresponding to the proton-electron interaction is given by
\begin{equation}
u_\mathrm{Coul}^{pe}(r)=-e^2\int \! d^3r' \, \rho_e(r')\frac{1}{|r-r'|}.
\end{equation}
Assuming that the electrons are uniformly distributed inside the cell, with the
density  $\rho_e=3Z/(4\pi R_{WS}^3)$, one gets
\begin{equation}
u_\mathrm{Coul}^{pe}(r)
=-2\pi e^2\rho_e\left(R_{WS}^2-\frac{1}{3}r^2\right) =
\frac{Ze^2}{2R_{WS}}\left(
\left(\frac{r}{R_{WS}}\right)^2-3\right)
\end{equation}
It can be seen that inside the WS cell the contribution of the
proton-electron interaction to the mean field is quadratic in the
radial coordinate.

\subsection{The electron and the lattice energies}

In the inner-crust the electrons are ultra-relativistic. Their
kinetic energy is given by the expression~\cite{landau}
\begin{equation}
T_e=Zm_ec^2 \left\{\frac{3}{8x^3}\left[x\left(1+2x^2\right)
\sqrt{1+x^2}-\ln\left(x+\sqrt{1+x^2}\right)\right]-1\right\},
\label{eq:Te}
\end{equation}
where $x$ is the relativistic parameter defined as $x=\hbar k_{Fe}/(m_ec^2)$.
In the ultra-relativistic regime $x\gg 1$.

The lattice energy is generated by the electron-proton and electron-electron
Coulomb interactions. The first one is given by
\begin{equation}
E_\mathrm{Coul}^{pe}=-\int \! d^3rd^3r' \,
\rho_p(r)\frac{e^2}{|r-r'|}\rho_e(r') =-\frac{3}{2}\frac{Z N_e
e^2}{R_{WS}}+2\pi\frac{e^2N_e}{R_{WS}^3}\int \! dr \, \rho_p(r)r^4
\; , \label{eq:Epe}
\end{equation}
where the last two terms on the right hand side are obtained assuming
that the
electron density $\rho_e$ is constant in the cell.

The electron-electron Coulomb energy is given by
\begin{equation}
E_\mathrm{Coul}^{ee}=
\frac{1}{2}\int \! d^3rd^3r' \, \rho_e(r)\frac{e^2}{|r-r'|} \rho_e(r')
-\frac{3}{4}\left(\frac{3}{\pi}\right)^{1/3} e^2 \int \! d^3 r \, \rho_e^{4/3}(r)
\end{equation}
where the  second term is the contribution
of the exchange term evaluated in the Slater approximation.
For a constant electron density one gets
\begin{equation}
E_\mathrm{Coul}^{ee}=\frac{3}{5}\frac{N_e^2e^2}{R_{WS}}\left(1
-\frac{5}{4}\left(\frac{3}{2\pi}\right)^{2/3}\frac{1}{N_e^{2/3}}\right).
\end{equation}

The Coulomb energy corresponding to the proton-proton interaction is calculated
within the mean field approach in a standard way, including the contribution of the
exchange term evaluated in the Slater approximation.

\section{Beta equilibrium condition}

Beta equilibrium condition is satisfied if $\delta \mu=0$ where
\begin{equation}
\delta \mu=m_ec^2+\mu_e+m_pc^2+\mu_p-m_nc^2-\mu_n \; .
\label{eq:beta}
\end{equation}

The chemical potential of the electrons can be written as
\begin{equation}
\mu_e=\sqrt{(\hbar c k_e)^2+(m_ec^2)^2}-m_ec^2+\mu_I^{ee}+\mu_I^{ep} ,
\label{eq:mue1}
\end{equation}
where $\mu_I^{ee}$ and $\mu_I^{ep}$ are the contributions coming from the
electron-electron and electron-proton interaction. They are given by:
\begin{equation}
\mu_I^{ee}=\frac{dE_\mathrm{Coul}^{ee}}{dN_e}
\end{equation}
\begin{equation}
\mu_I^{ep}=\frac{dE_\mathrm{Coul}^{ep}}{dN_e}
\end{equation}

The chemical potentials of the neutrons and protons are extracted
from the HFB calculations. The contribution of the proton-electron
interaction to the chemical potential of the protons is included
through the proton-electon mean field (14).

As seen in Eq. (12), the proton 
mean field includes also the contribution of the proton-electron 
interaction   

The beta equilibrium condition can be satisfied exactly when the
chemical potential of the neutrons, determined by the nonlocalized
neutrons, is a continuous
variable. In the calculations done here the neutron spectrum is
discretized due to the boundary conditions imposed at the border
of the cells (it is worth mentioning that this discretization has
nothing to do with the discrete structure of the neutron spectrum
generated by the symmetry of the crystal lattice). Consequently
the beta equilibrium
condition is satisfied only approximatively. In practice, we
consider that the beta equilibrium condition is found when by
changing the N/Z ratio the value of $\delta \mu$ is changing the
sign. Then, from the two N/Z configurations for which $\delta \mu$
is changing the sign we keep the one which has the smaller binding
energy. It is worth stressing that the beta equilibrium condition
depends, through the discretization of the neutron spectrum, on
the type of boundary conditions imposed  at the border of the
cells. How the type of boundary conditions could influence the
structure of the cells is discussed in Ref~\cite{baldo2006}.

\section{Results: the structure of the Wigner-Seitz cells}

Within the framework presented in the previous sections we have determined
the properties of the WS cells, i.e. the N/Z ratio and the radius
of the cells. The calculations have been done for the set of baryonic
densities shown in Table~\ref{tab:NV}. To find the structure of the cell at
a given density we have considered all the configurations with the even
number of protons between 12 and 60. For each number of protons we modified
the radius of the cell with a step of 0.2 fm, keeping the same total density,
until the number of neutrons included in the cell satisfies  with the best
accuracy the beta equilibrium condition. The most probable configuration
at a given density is finally taken as the one with the lowest binding energy.

First we have determined the structure of the WS cells
in the HF approximation, i.e., neglecting the contribution of
pairing correlations. The results are given in Table~\ref{tab:hf}.
Compared to previous calculations \cite{negele,baldo}
we find that the cells have a smaller number of protons.
Table II shows also that the number of protons are not anymore
equal to a magic or a semi-magic number as in Ref.\cite{negele} (see
Table I).

\begin{figure}[ht]
\begin{center}
\includegraphics[scale=1]{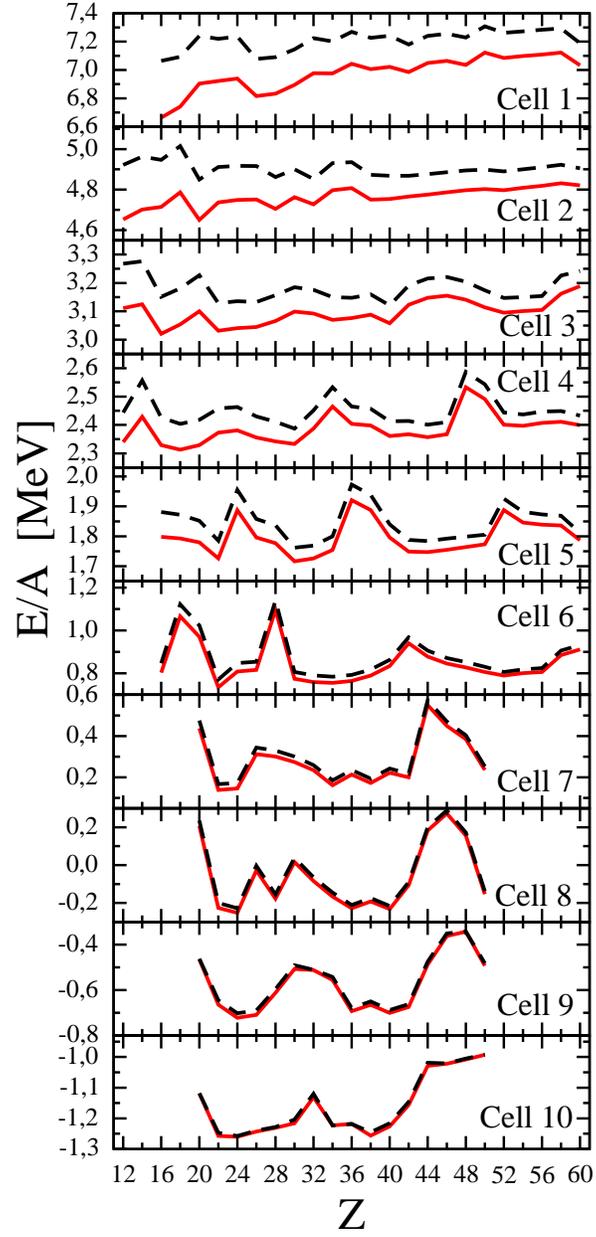}
\caption{(color online) HF energies per particle versus the
proton number Z (full lines). With the dashed lines are
represented the HF results after the finite size corrections.}
\label{fig:ehf}
\end{center}
\end{figure}

\begin{figure}[ht]
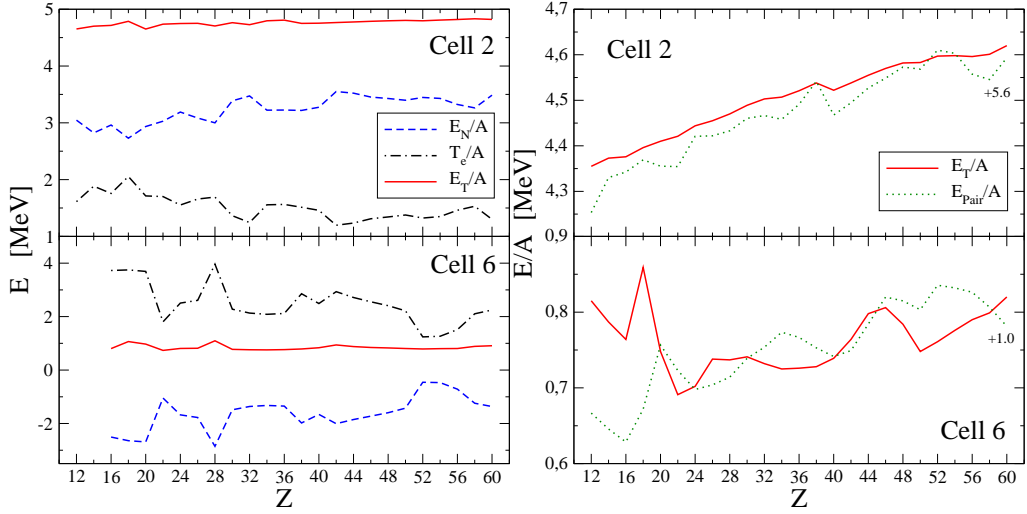

\begin{center}
\includegraphics[scale=0.4]{fig_ehf1.eps}
\includegraphics[scale=0.4]{fig_ehf2.eps}
\caption{(color online) The different contributions to the total
energy in the cells 2 and 6 for the HF (left pannel) and HFB
(right pannel) calculations. Are shown: the total energy (solid line),
the nuclear energy (dashed line), the kinetic energy of the electrons
(dashed-dotted line), and the pairing energy for the ISS pairing
interaction (dotted line). The pairing energies are shifted up
as indicated in the figure.}
\label{fig:ehf2}
\end{center}
\end{figure}

 To understand better the results of Table II, in Fig.~\ref{fig:ehf}
we show the evolution of the binding energies per nucleon, calculated
at beta equilibrium and at constant density, with respect to the
proton number $Z$. The most probable configuration corresponds to
the number of protons for which the binding energy has the lowest
value. From Fig.~\ref{fig:ehf} it can be seen that in the cells 1
and 2 there is a continuous decrease of the binding energy for the
lowest values of $Z$. Thus, for these cells the HF calculations
cannot predict a well-defined cell structure. From
Fig.~\ref{fig:ehf} it can be also seen that even for the cells in
which one can identify a configuration with the lowest binding
energy, the difference
 between this energy and the energy of other local minima is very small, of the
 order of 10 keV. The weak dependence of the binding energy on $Z$ observed in
Fig.~\ref{fig:ehf} is caused by the almost exact compensation
between the nuclear energy and the electron kinetic energy. This
can be clearly seen in  Fig.~\ref{fig:ehf2} where are shown, for
the cells 2 and 6, the contributions to the total energy coming
from the nuclear energy (dashed line) and the kinetic
energy of the electrons (dashed-dotted line). It can be  noticed that the local
minima of nuclear binding energy at $Z=20$ and $Z=28$ are washed
out by the kinetic energy of the electrons. The competition between the
nuclear and Coulomb interaction, specific to the so-called
frustrated systems, it is the reason why the structure of the
WS cells it is not necessarily
determined by the nuclear interaction and the associated nuclear
shell effects.

\begin{figure}[ht]
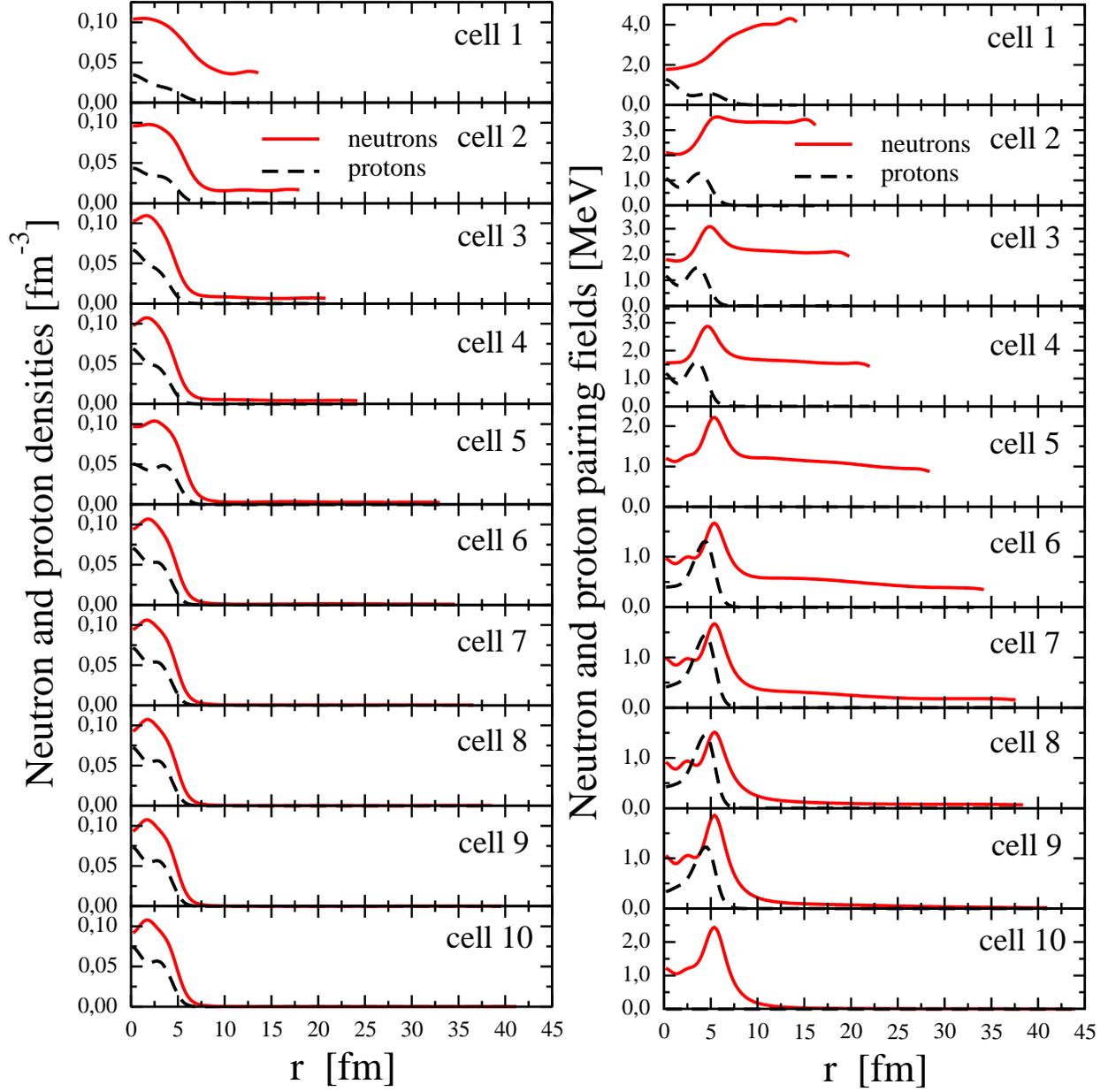

\begin{center}
\includegraphics[scale=1]{fig_den1bis.eps}
\includegraphics[scale=1]{fig_den2.eps}
\caption{(color online) The radial profiles of densities
(left) and  pairing fields (right) for neutrons (full lines) and
protons (dashed lines)  . The densities
correspond to the HF calculations while the pairing fields
to the HFB calculations with the pairing force ISS.} \label{fig:den}
\end{center}
\end{figure}

A necessary condition for the validity of the WS approximation is
the appearance 
in the neutron density of a well-defined plateau before the edge
of the cell. From Fig.~\ref{fig:den} (left pannel) it can be
observed that this condition is hardly fulfilled for the cell 1,
reasonably well for the cell 2 and better for the other cells.

We shall now discuss the effect of pairing correlations on the
structure of the WS cells. To study the influence of pairing
correlations we have performed HFB calculations with the three
pairing interactions introduced in Section 3A. How the pairing
correlations are distributed in the cells is illustrated in Fig.4
(right pannel) which shows the pairing fields of neutrons and
protons for the force ISS. As expected, the pairing field profile
has a non-uniform distribution which could be traced back to the
density dependence of the pairing force \cite{sandulescu1}. The
proton pairing field stays localized inside the nuclear cluster
since a drip out of protons is not observed in our calculations.
It can be noticed that for the cells 5 and 10, with the proton
numbers  $Z=20$ and $Z=28$ (see Table~\ref{tab:hfb}), the proton
fields are zero. This indicates that in the nuclear clusters
corresponding to these cells the proton numbers $Z=20,28$ behave
as magic numbers, as in atomic nuclei.

 The dependence of  pairing energy on Z is illustrated in Fig. 3 (right pannel)
for the cells number 2 and 6. One observes that in average the absolute value
of the pairing energy is decreasing with Z, which shows that the dominant
contribution to pairing comes from the nonlocalized neutrons 
(notice that for a cell the HFB
 calculations with various Z are done for a fixed total density).

\begin{figure}[ht]
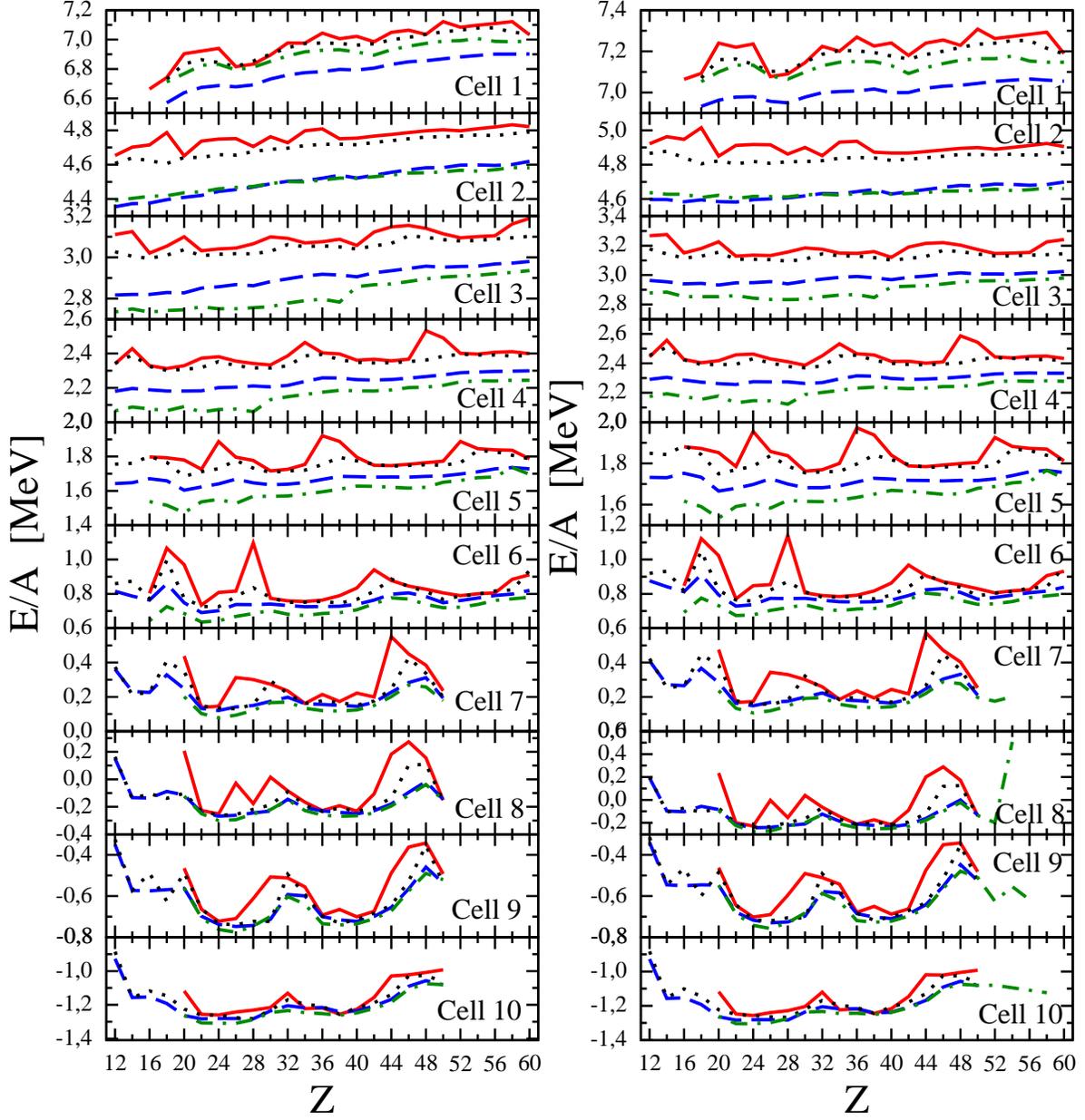

\begin{center}
\includegraphics[scale=1]{fig_ehfb.eps}
\includegraphics[scale=1]{fig_ehfbc.eps}
\caption{(color online) The HFB energies per particle as function of
proton number for the pairing forces ISW (dotted line),
ISS (dashed line) and IVS (dashed-dotted line). The solid lines
represent the HF results. In the left pannel are shown the results
obtained including the finite size corrections.}
\label{fig:ehfb}
\end{center}
\end{figure}

The structure of the WS cells obtained in the HFB approach
is given in Table~\ref{tab:hfb} while in  Fig.~\ref{fig:ehfb} it is
shown the dependence of the binding energies, at beta equilibrium,
on protons number.  From Fig.~\ref{fig:ehfb} we observe that in 
the cell 1 the binding energy does not converge to a minimum 
before Z=12. For the cells 2-4 a minimum can be found for the
ISW and/or IVS forces but this minimum is very close to the
value of binding energy at Z=12. Therefore the structure of the 
cells 2-4 is ambiguous. The situation is different in
the cells 5-10 where the binding energies converge to absolute
minima located before Z=12. Thus for these cells the 
structure can be well-defined by the present HFB calculations.

Comparing  Table~\ref{tab:hfb} and Table~\ref{tab:hf}
it can be observed that for the cells 6-9  the numbers of protons
in the HF and HFB calculations differ by about 2 units. The largest difference, of 10 units, appears
for the cell 5. However, as seen in  Fig.~\ref{fig:ehfb}, the HF minimum at Z=30 is
in fact very close to the local minimum at Z=22. A similar situation
can be noticed in cell 10 for the HF minima at Z=24 and Z=28.
In conclusion, these calculations indicate that the pairing does
not change much the structure of the low density cells 5-10. 
This could be also observed from the fact that in these cells
the intensity of pairing force has only marginal effects on the
proton and neutron numbers.

Let us now discuss more in detail what happens in the high density
region for the configurations with small Z and small cells radii.
When the radius of the cell becomes too small the boundary conditions
imposed at the cell border through the WS approximation generate an
artificial large distance between the energy levels of the nonlocalized
neutrons. Consequently, the binding energy of the neutron gas is significantly
underestimated. An estimation  of how large could be the errors in
the binding energy induced by the WS approximation can be obtained
from the quantity
\begin{equation}
f(\rho_{n},R_{WS}) \equiv
B_{inf.}(\rho_{n}) - B_{WS-inf.}(\rho_{n},R_{WS}) \; ,
\end{equation}
where the first term is the binding energy per neutron for infinite neutron
matter of density $\rho_n$ and the second term is the binding energy of neutron
matter with the same density calculated inside the cell of radius $R_{WS}$
and employing  the same boundary conditions as in HF or HFB calculations.
In Ref.~\cite{margueron2007} it was proposed for the finte size energy 
correction, Eq. 26, the following parametrisation
\begin{equation}
f(\rho_{n_g},R_{WS}) = 89.05 (\rho_{n_g}/\rho_0)^{0.1425}
R_{WS}^{-2} \; ,
\end{equation}
where $\rho_{n_g}$ is the average density of neutrons in the gas
region extracted from a calculation in which the cell contains
both the nuclear cluster and the nonlocalized neutrons while
$\rho_0$ is the nuclear matter saturation density.

How the energy corrections described by Eq.(25) influences the HF (HFB)
results can be seen in Tables~III (Tables-IV) and Fig.~2 (Fig.~5).  As
expected, the influence of the corrections is more important for
the cells 1-5, in which the neutron gas has a higher density, and
for those configurations corresponding to small cell radii. For
the cell 1 the binding energy after the correction is still
decreasing for the smallest Z values, which means that the structure
of this cell remains uncertain. The structure of the cells 2-4 can 
be now determined for all pairing forces. However, as seen in
Fig. 5, for these cells the absolute minima are still very close to the 
binding energies at Z=12 which shows that the structure of these cells
remains ambiguous even after the energy correction.

\section {Summary and conclusions}

 In this paper we have examined the influence of pairing correlations on the
structure of inner crust of neutron stars. The study was done for the region
of the inner crust which is supposed to be formed by a lattice
of spherical clusters. The lattice was treated as a set of independent
cells described in the Wigner-Seitz approximation. To determine the structure
of a cell we have used the nuclear binding energy given by the HFB approach.
For the HFB calculations we have considered a particle-hole interaction
of Skyrme type (SLy4) while as particle-particle interaction we have used
three zero range density-dependent pairing forces of various intensities.
The calculations show that the pairing correlations have a weak influence
on the structure of WS cells. 

 For the cells with high density and small radii the binding energies do
not converge to a minimum when the proton number has small values.
We believe
that the reason for that is the failure of the WS approximation
when the cell radius is too small. For a small radius of the cell
the average distance between the energy levels of the nonlocalised
neutrons becomes artificially large which cause an
underestimation of the binding energy. To correct this drawback we
have used an empirical expression based on the comparison
between the binding energy of neutrons calculated in infinite
matter geometry and in a spherical cell \cite{margueron2007}. We
found that the corrections to the binding energies are significant
for the high density cells with small proton numbers. This show
that the WS approximation is not accurately enough for predicting
the structure of the high density region of the inner crust.
\vskip 0.5cm {\bf Acknowledgements}

This work was supported by the European Science Foundation through the
project "New Physics of Compact Stats", by the Romanian Ministry of Research
and Education through the grant Idei nr. 270 and by the French-Romanian
collaboration IN2P3-IFIN.

\newpage

\begin{table}[ht]
\setlength{\tabcolsep}{.12in}
\renewcommand{\arraystretch}{0.9}
\begin{tabular}{cccccc}
\toprule
   $N_{cell}$ & $\rho$  & $N$ & $Z$ & $R_{WS}$ \\
    & [g cm$^{-3}$] & & & [fm] \\
\colrule
  1 & $7.9\ 10^{13}$ & 1460 &40  &20  \\
  2 & $3.4\ 10^{13}$ & 1750 &50  &28 \\
  3 & $1.5\ 10^{13}$ & 1300 &50  &33  \\
  4 & $9.6\ 10^{12}$ & 1050 &50  &36  \\
  5 & $6.2\ 10^{12}$ & 900 &50  &39  \\
  6 & $2.6\ 10^{12}$ & 460 &40  &42  \\
  7 & $1.5\ 10^{12}$ & 280 &40  &44  \\
  8 & $1.0\ 10^{12}$ & 210 &40  &46  \\
  9 & $6.6\ 10^{11}$ & 160 &40  &49  \\
 10 & $4.6\ 10^{11}$ & 140 &40  &54  \\
\botrule
\end{tabular}
\caption{The structure of the Wigner-Seitz cells
determined in Ref.\cite{negele}. $\rho$ is the baryon
density, N and Z are the numbers of
neutrons and protons while $R_{WS}$ is the radius of the cell.
Compared to Ref. \cite{negele} here it is not shown the cell with the
highest density located at the interface with the pasta phase. }
\label{tab:NV}
\end{table}

\begin{table}[ht]
\setlength{\tabcolsep}{.1in}
\renewcommand{\arraystretch}{0.9}
\begin{tabular}{cccccccccc}
\toprule
   $N_{cell}$ & $\rho$ & $N$ & $Z$ & $R_{WS}$ & $E/A$ & $E_N/A$ & $T_e/A$ & $\mu_n$ & $\mu_p$\\
    & [g cm$^{-3}$] & & & [fm] & [MeV] & [MeV] & [MeV] & [MeV] & [MeV] \\
\colrule
 $ 3$ & $1.5\ 10^{13}$ & $318$ & $16$ & $20.8$ & $ 3.021$ & $ 1.409$ & $ 1.622$ & $ 4.591$ & $-38.03$ \\
 $ 4$ & $9.6\ 10^{12}$ & $322$ & $18$ & $24.2$ & $ 2.313$ & $ 0.724$ & $ 1.600$ & $ 4.169$ & $-35.94$ \\
 $ 5$ & $6.2\ 10^{12}$ & $528$ & $30$ & $33.0$ & $ 1.716$ & $ 0.310$ & $ 1.409$ & $ 2.996$ & $-31.22$ \\
 $ 6$ & $2.6\ 10^{12}$ & $252$ & $22$ & $34.6$ & $ 0.735$ & $-1.047$ & $ 1.805$ & $ 1.886$ & $-27.49$ \\
 $ 7$ & $1.5\ 10^{12}$ & $158$ & $22$ & $36.6$ & $ 0.139$ & $-2.413$ & $ 2.590$ & $ 1.025$ & $-26.90$ \\
 $ 8$ & $1.0\ 10^{12}$ & $120$ & $24$ & $38.6$ & $-0.252$ & $-3.649$ & $ 3.447$ & $ 0.773$ & $-26.27$ \\
 $ 9$ & $6.6\ 10^{11}$ & $ 80$ & $24$ & $39.8$ & $-0.722$ & $-5.294$ & $ 4.643$ & $ 0.294$ & $-26.07$ \\
 $10$ & $4.6\ 10^{11}$ & $ 58$ & $24$ & $41.4$ & $-1.260$ & $-6.812$ & $ 5.648$ & $-0.312$ & $-26.06$ \\
\botrule
\end{tabular}
\caption{The structure of Wigner-Seitz cells obtained in the HF
approximation with the force SLy4. E/A, E$_N$/A and T$_e$/A
are, respectively, the total energy, the nuclear energy and
the electron kinetic energy per baryon while $\mu_n$ and $\mu_p$
are  the neutron chemical potential and the proton chemical
potential. The other quantities are the same as in Table I.
The structure of the cells 1-2 it is not shown because it
is not well-defined  by the present HF calculations.} \label{tab:hf}
\end{table}

\begin{table}[ht]
\renewcommand{\arraystretch}{0.9}
\begin{tabular}{ccccccccccccccccc}
\toprule
$N_{cell}$ & \multicolumn{3}{c}{$N$} & \multicolumn{3}{c}{$Z$} & \multicolumn{3}{c}{$R_{WS}$  [fm] } & \multicolumn{3}{c}{$E/A$  [MeV]} & \multicolumn{3}{c}{$\mu_n$  [MeV]} & $\mu_p$  [MeV]\\
  & ISW & ISS & IVS & ISW & ISS & IVS & ISW & ISS & IVS & ISW & ISS & IVS & ISW & ISS & IVS & ISW\\
\colrule
  2 & $476$ &       &       & $18$ &       &      & $18.0$ &        &        & $ 4.607$ &          &          & $ 6.915$ &          &          & $-46.95$\\
  3 & $368$ &       & $378$ & $16$ &       & $16$ & $21.8$ &        & $20.0$ & $ 2.995$ &          & $ 2.734$ & $ 4.864$ &          & $4.395$  & $-37.76$\\
  4 & $330$ &       & $460$ & $18$ &       & $22$ & $24.4$ &        & $27.2$ & $ 2.302$ &          & $ 2.059$ & $ 3.820$ &          & $3.431$  & $-35.71$\\
  5 & $320$ & $336$ & $344$ & $20$ & $ 20$ & $20$ & $28.0$ & $28.4$ & $28.6$ & $ 1.685$ & $ 1.604$ & $ 1.473$ & $ 2.974$ & $ 2.828$ & $2.629$  & $-32.57$\\
  6 & $300$ & $242$ & $238$ & $24$ & $ 22$ & $22$ & $36.6$ & $34.2$ & $34.0$ & $ 0.725$ & $ 0.691$ & $ 0.636$ & $ 1.631$ & $ 1.678$ & $1.546$  & $-32.45$\\
  7 & $202$ & $170$ & $174$ & $26$ & $ 24$ & $24$ & $39.6$ & $37.6$ & $37.8$ & $ 0.131$ & $ 0.121$ & $ 0.078$ & $ 1.152$ & $ 1.052$ & $0.988$  & $-25.98$\\
  8 & $120$ & $118$ & $120$ & $24$ & $ 24$ & $24$ & $38.6$ & $38.4$ & $38.6$ & $-0.262$ & $-0.268$ & $-0.300$ & $ 0.607$ & $ 0.696$ & $0.654$  & $-28.56$\\
  9 & $ 92$ & $ 90$ & $ 94$ & $26$ & $ 26$ & $26$ & $41.4$ & $41.4$ & $41.6$ & $-0.739$ & $-0.748$ & $-0.776$ & $ 0.402$ & $ 0.384$ & $0.403$  & $-25.51$\\
 10&  $ 84$ & $ 72$ & $ 64$ & $28$ & $ 28$ & $26$ & $46.0$ & $44.2$ & $42.6$ & $-1.282$ & $-1.283$ & $-1.310$ & $ 0.238$ & $ 0.167$ & $0.092$  & $-23.29$\\
\botrule
\end{tabular}
\caption{The structure of Wigner-Seitz cells obtained in the HFB
approximation. The results corresponds to the isoscalar
weak (ISW), isoscalar strong (ISS) and isovector-isoscalar
(IVS) pairing forces. The displayed quantities are the
same as in Table II. In the table are shown only the structures of
the cells which could be well-defined by the HFB calculations.}
\label{tab:hfb}
\end{table}

\begin{table}[ht]
\setlength{\tabcolsep}{.12in}
\renewcommand{\arraystretch}{0.9}
\begin{tabular}{ccccccccc}
\toprule
   $N_{cell}$ & $N$ & $Z$ & $R_{WS}$ & $E/A$ & $E_N/A$ & $T_e/A$ & $\mu_n$ & $\mu_p$\\
  & & & [fm] & [MeV] & [MeV] & [MeV] & [MeV] & [MeV] \\
\colrule
 $ 2$ & $474$ & $20$ & $18.0$ & $ 4.850$ & $ 2.935$ & $ 1.713$ & $ 6.046$ & $-47.49$ \\
 $ 3$ & $980$ & $40$ & $30.2$ & $ 3.121$ & $ 1.806$ & $ 1.241$ & $ 4.651$ & $-37.01$ \\
 $ 4$ & $726$ & $30$ & $31.6$ & $ 2.387$ & $ 1.246$ & $ 1.088$ & $ 3.953$ & $-31.67$ \\
 $ 5$ & $538$ & $30$ & $33.2$ & $ 1.762$ & $ 0.344$ & $ 1.376$ & $ 3.080$ & $-31.17$ \\
 $ 6$ & $252$ & $22$ & $34.6$ & $ 0.771$ & $-1.047$ & $ 1.805$ & $ 1.886$ & $-27.49$ \\
 $ 7$ & $158$ & $22$ & $36.6$ & $ 0.167$ & $-2.413$ & $ 2.590$ & $ 1.025$ & $-26.90$ \\
 $ 8$ & $120$ & $24$ & $38.6$ & $-0.228$ & $-3.649$ & $ 3.447$ & $ 0.773$ & $-26.27$ \\
 $ 9$ & $ 80$ & $24$ & $39.8$ & $-0.702$ & $-5.294$ & $ 4.643$ & $ 0.294$ & $-26.07$ \\
 $10$ & $ 58$ & $24$ & $41.4$ & $-1.257$ & $-6.812$ & $ 5.648$ & $-0.312$ & $-26.06$ \\
\botrule
\end{tabular}
\caption{The structure of the Wigner-Seitz cells obtained in the HF
approximation including the finite size corrections
(see text for details). The displayed quantities are the same as in Table II.} \label{tab:hfc}
\end{table}

\begin{table}[ht]
\renewcommand{\arraystretch}{0.9}
\begin{tabular}{cccccccccccccccccc}
\toprule $N_{cell}$ & \multicolumn{3}{c}{$N$} &
\multicolumn{3}{c}{$Z$} & \multicolumn{3}{c}{$R_{WS}$  [fm] } & \multicolumn{3}{c}{$E/A$  [MeV]} & \multicolumn{3}{c}{$\mu_n$ [MeV]} & $\mu_p$  [MeV]\\
  & ISW & ISS & IVS & ISW & ISS & IVS & ISW & ISS & IVS & ISW & ISS & IVS & ISW & ISS & IVS & ISW \\
\colrule
  2 & 656 & 676 & 656 & 22 & 22 & 22 & $20.0$ & $20.2$ & $20.0$ & 4.804  & 4.582  & 4.603  & 6.958  & 6.785  & 6.785  & $-45.60$ \\
  3 & 718 & 454 & 734 & 28 & 20 & 28 & $27.2$ & $23.4$ & $27.4$ & 3.095  & 2.932  & 2.833  & 4.933  & 4.492  & 4.472  & $-35.93$ \\
  4 & 712 & 460 & 594 & 30 & 22 & 28 & $31.4$ & $27.2$ & $29.6$ & 2.370  & 2.256  & 2.122  & 3.714  & 3.607  & 3.446  & $-37.67$ \\
  5 & 320 & 336 & 344 & 20 & 20 & 20 & $28.0$ & $28.4$ & $28.6$ & 1.749  & 1.666  & 1.534  & 2.974  & 2.828  & 2.629  & $-32.57$ \\
  6 & 300 & 242 & 238 & 24 & 22 & 22 & $36.6$ & $34.2$ & $34.0$ & 0.758  & 0.729  & 0.674  & 1.631  & 1.678  & 1.546  & $-32.45$ \\
  7 & 202 & 170 & 174 & 26 & 24 & 24 & $39.6$ & $37.6$ & $37.8$ & 0.156  & 0.148  & 0.106  & 1.152  & 1.052  & 0.988  & $-25.98$ \\
  8 & 120 & 118 & 120 & 24 & 24 & 24 & $38.6$ & $38.4$ & $38.6$ & -0.238 & -0.244 & -0.276 & 0.607  & 0.696  & 0.654  & $-28.56$ \\
  9 &  92 &  90 &  94 & 26 & 26 & 26 & $41.4$ & $41.2$ & $41.6$ & -0.721 & -0.730 & -0.758 & 0.402  & 0.384  & 0.403  & $-25.51$ \\
 10&   84 &  72 &  64 & 28 & 28 & 26 & $46.0$ & $44.2$ & $42.6$ & -1.268 & -1.283 & -1.310 & 0.238  & 0.167  & 0.092  & $-23.29$ \\
\botrule
\end{tabular}
\caption{The structure of Wigner-Seitz cells obtained in the HFB
approximation including the finite size corrections. The displayed quantities
are the same as in Table III.}
\label{tab:hfbc}
\end{table}

\end{document}